\documentclass[a4paper]{jpconf}

\usepackage[dvips]{graphicx}
\usepackage{latexsym,amssymb,amsmath,epsfig,bm,times,psfrag,subfigure}
\usepackage{color}
\usepackage[bookmarksnumbered,bookmarksopen,colorlinks,citecolor=red,linkcolor=blue]{hyperref}
\renewcommand\r{\rho}

\renewcommand\t{\tau}

\renewcommand\o{\omega}
\renewcommand\e{\epsilon}
\newcommand\g{\gamma}

\newcommand\m{\mu}
\newcommand\n{\nu}

\newcommand\p{\pi}

\newcommand\s{\sigma}

\newcommand\w{\eta}
\newcommand\ve{\varepsilon}

\renewcommand\L{\Lambda}

\newcommand\F{\Phi}

\newcommand{\fig}[1]{Fig.~\ref{#1}}

\newcommand\ls{\left[}
\newcommand\rs{\right]}

\newcommand{\lan}{\langle}
\newcommand{\ran}{\rangle}

\renewcommand\pt{\partial}

\newcommand{\bp}{{\bm p}}

\newcommand{\bv}{{\bm v}}

\renewcommand{\vec}{\boldsymbol}
\newcommand{\be}{\begin{equation}}
\newcommand{\ee}{\end{equation}}
\newcommand{\bear}{\begin{eqnarray}}
\newcommand{\eear}{\end{eqnarray}}
\newcommand{\ba}{\begin{array}}
\newcommand{\ea}{\end{array}}

\begin{document}
\title{Event-by-event generation of vorticity in heavy-ion collisions}
\author{Wei-Tian Deng}
\address{School of physics, Huazhong University of Science and Technology, Wuhan 430074, China}
\author{Xu-Guang Huang}
\address{Physics Department and Center for Particle Physics and Field Theory, Fudan University, Shanghai 200433, China}
\address{Department of Physics, Brookhaven National Laboratory, Upton, New York 11973-5000, USA}
\ead{huangxuguang@fudan.edu.cn}

\begin{abstract}
In a noncentral heavy-ion collision, the two colliding nuclei have finite angular momentum in the direction perpendicular to the reaction plane. After the collision, a fraction of the total angular momentum is retained in the produced hot quark-gluon matter and is manifested in the form of fluid shear. Such fluid shear creates finite flow vorticity. We study some features of such generated vorticity, including its strength, beam energy dependence, centrality dependence, and spatial distribution.
\end{abstract}

\section{Introduction}
Relativistic heavy-ion collisions provide us the environments in which we can study the strongly interacting matter under unusual conditions, like extremely high temperature~\cite{Wang:2016opj} and extremely strong magnetic field~\cite{Huang:2015oca,Hattori:2016emy}. Recently, it was realized that noncentral heavy-ion collisions can also generate finite flow voriticty and thus provide us a chance to study quark-gluon matter under local rotation~\cite{Csernai:2013bqa,Becattini:2015ska,Jiang:2016woz,Deng:2016gyh}. The mechanism of the generation of finite vorticity is simple. In a noncentral heavy-ion collision, the two colliding nuclei have a finite angular momentum in the direction perpendicular to the reaction plane, $J_0\sim Ab\sqrt{s}/2$, with $A$ the number of nucleons in one nucleus, $b$ the impact parameter, and $\sqrt{s}$ the center-of-mass energy of a pair of colliding nucleons. After the collision, a fraction of the total angular momentum is retained in the produced partonic matter which we will call the quark-gluon plasma (QGP). This fraction of angular momentum manifests itself as a shear of the longitudinal momentum density and thus nonzero local vorticity arises. In hydrodynamics, the vorticity represents the local angular velocity, and its existence in heavy-ion collisions may induce a number of intriguing phenomena, for example, the chiral vortical effect ~\cite{Kharzeev:2015znc}, the chiral vortical wave~\cite{Jiang:2015cva}, and the global polarization of quarks and $\L$ baryons~\cite{Liang:2004ph,Huang:2011ru,Becattini:2013fla,Pang:2016igs,Fang:2016vpj,Lisa}.

We present detailed numerical study of the event-by-event generation of vorticity in heavy-ion collisions by using the HIJING event generator~\cite{Deng:2016gyh}.

\section{Numerical setups}
The coordinate system is setup as the following. We choose the $z$ axis to be along the beam direction of the projectile, $x$ axis to be along the impact parameter ${\vec b}$ which points from the target to the projectile, and $y$ axis to be perpendicular to the reaction plane. The origin of the time axis, $t=0$, is set to the time when the two colliding nuclei overlap maximally.

To calculate the vorticity field, we need to know the velocity field $\vec v(x)$. In relativistic system, the definition of the velocity is not unique. We define two types of velocity field, $\vec v_1$ and $\vec v_2$, which represent the velocity of the particle flow and the velocity of the energy flow, respectively,
\begin{eqnarray}
\label{defv1}
v^a_1(x)&=&\frac{1}{\sum_i \F(x,x_i)}\sum_i\frac{p^a_i}{p^0_i}\F(x,x_i),\\
\label{defv2}
v^a_2(x)&=&\frac{\sum_i p^a_i \F(x,x_i)}{\sum_i [p^0_i+(p_i^a)^2/p_i^0] \F(x,x_i)},
\end{eqnarray}
where $a=1, 2, 3$ is the spatial indices, $\bp_i$ and $p^0_i$ are the momentum and energy of the $i$th particle, and the summation is over all the particles. In our simulation, $p_i$ and $x_i$ in each event are generated by HIJING. The function $\F(x,x_i)$ is to smear a physical quantity, e.g., energy or momentum, carried by the $i$th particle at $x_i$ to another coordinate point $x$. Therefore, $\F(x,x_i)$ somehow represents the quantum wave packet of the $i$th particle whose functional form at $\t=\t_0$ is chosen to be a Gaussian~\cite{Pang:2012he},
\begin{eqnarray}
\label{smear2}
\F_{\rm G}(x,x_i)=\frac{K}{\t_0\sqrt{2\p\s_\w^2}2\p\s_r^2}\exp{\ls-\frac{(x-x_i)^2+(y-y_i)^2}{2\s_r^2}-\frac{(\w-\w_i)^2}{2\s_\w^2}\rs},
\end{eqnarray}
where $\s_r$ and $\s_\w$ are two width parameters and $K$ is a scale factor. The parameters are given in Refs.~\cite{Deng:2016gyh}.

Once the specific definition of the velocity field is given in numerical setup, the vorticity is calculated according to
\begin{eqnarray}
\label{define-vor1}
{\bm\o}_{1}&=&{\bm\nabla}\times\bv,\\
\label{define-vor2}
{\bm\o}_{2}&=&\g^2{\bm\nabla}\times\bv,
\end{eqnarray}
where $\vec\o_1$ is the usual nonrelativistic definition and $\vec\o_2$ is the spatial components of the relativistic definition, $\o^\m=\e^{\m\n\r\s}u_\n \pt_\r u_\s$ with $O(v^2)$ term omitted except for $\g$.

\section{Numerical results}
\begin{figure}[!htb]
\begin{center}
\includegraphics[width=6cm]{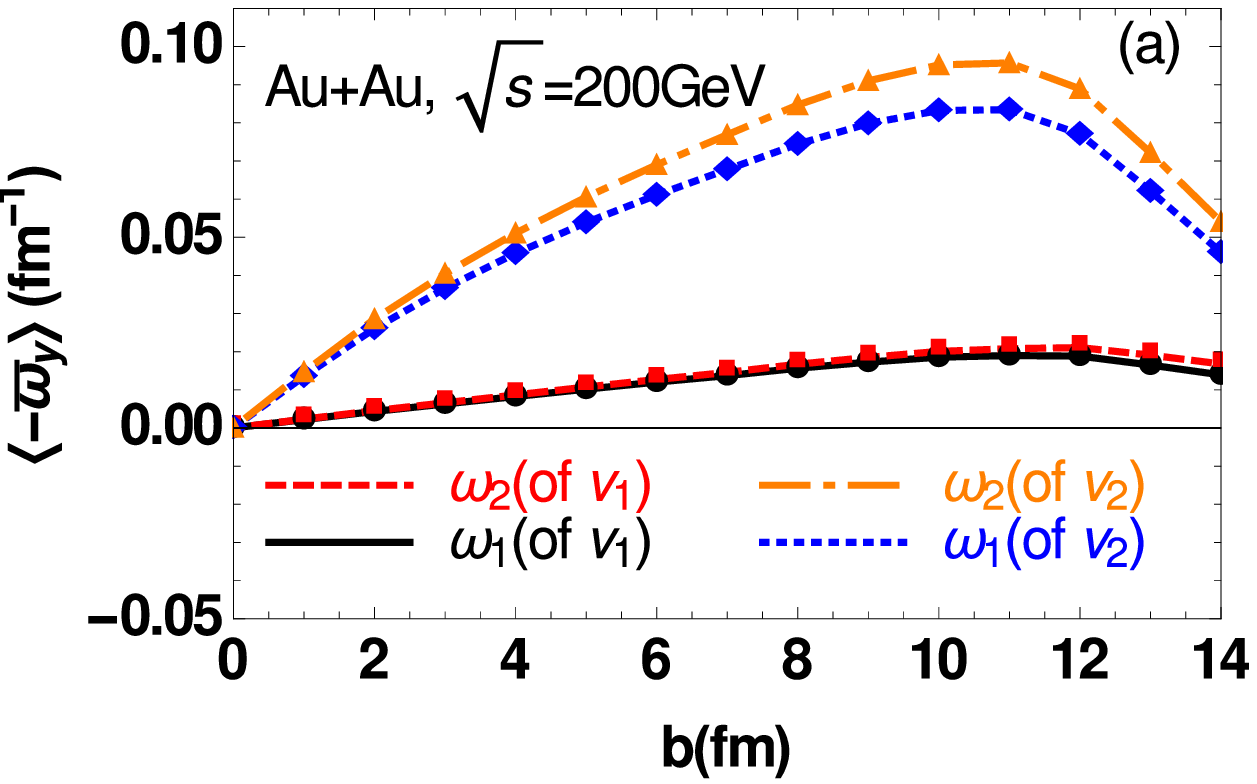}
\includegraphics[width=6cm]{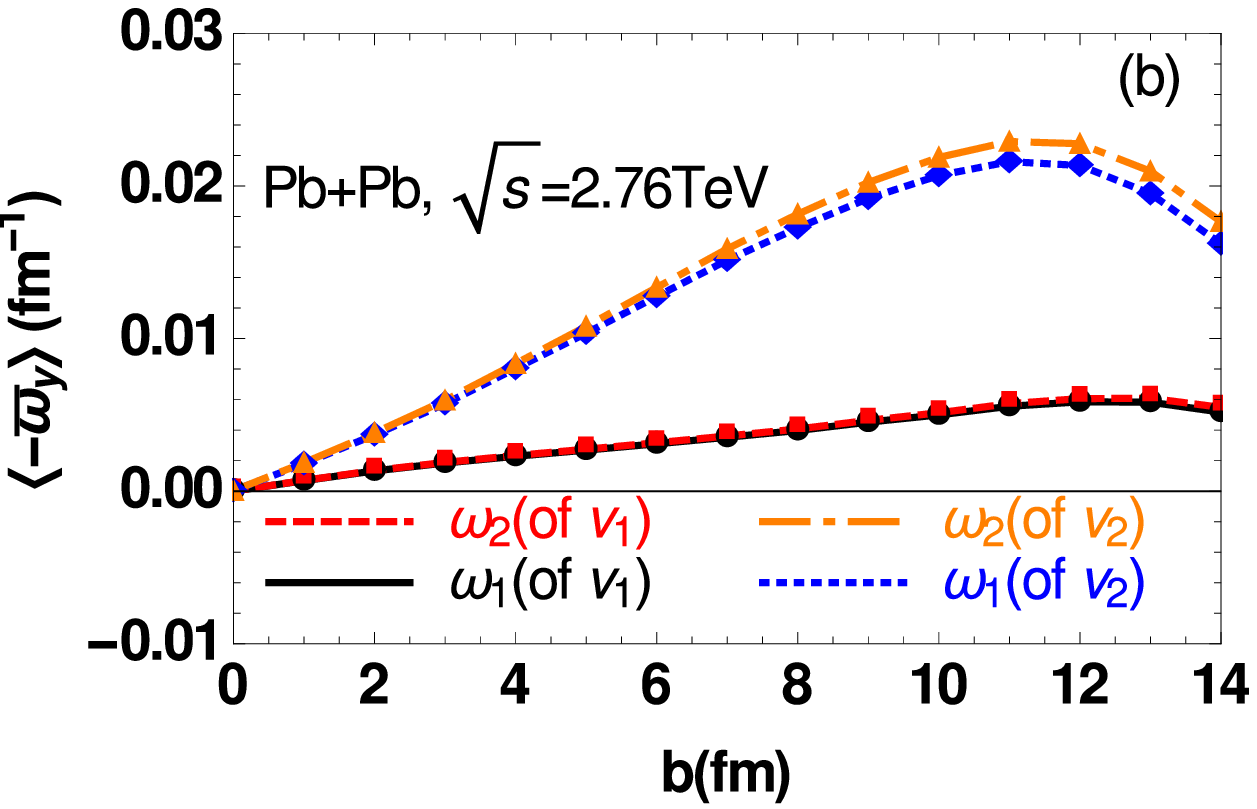}
\caption{The space-averaged vorticity at $\t=\t_0$ and $\w=0$ averaged over $10^5$ events for RHIC Au + Au collisions at $\sqrt{s}=200$ GeV (Left) and LHC Pb + Pb collisions at $\sqrt{s}=2.76$ TeV (Right).}
\label{fig-vor-b-200}
\end{center}
\end{figure}

In \fig{fig-vor-b-200}, we show the space-averaged $y$-component of the vorticities at $\t=\t_0$ and $\w=0$ averaged over $10^5$ events. The spatial average is performed by using number density as weight if the vorticity is computed by using the particle flow velocity $\bv_1$, or by using $\ve$ as weight if the vorticity is computed by using the energy flow velocity $\bv_2$. For $b\lesssim 2 R_A$ ($R_A$ is the radius of the nucleus), the vorticity increases with centrality. The magnitude of the vorticity at $\sqrt{s}=200$ GeV is big in the sense that it may be comparable to the magnetic field, $T^2\o\sim \m eB$ (for, say, temperature $T\sim 350$ MeV and baryon chemical potential $\m=10$ MeV).
 %%%%%%%%%%%%%%%%%%%%%%%%%%%%%%%%%%%%%%%%%%%%%%%%%%%%%%%%%%%%%%%%%%%%%%%
%%%%%%%%%%%%%%%%%%%%%%%%%%%%%%%%%%%%%%%%%%%%%%%%%%%%%%%%%%%%%%%%%%%%%%%

%%%%%%%%%%%%%%%%%%%%%%%%%%%%%%%%%%%%%%%%%%%%%%%%%%%%%%%%%%%%%%%%%%%%%%%
\begin{figure}[!htb]
\begin{center}
\includegraphics[width=6cm]{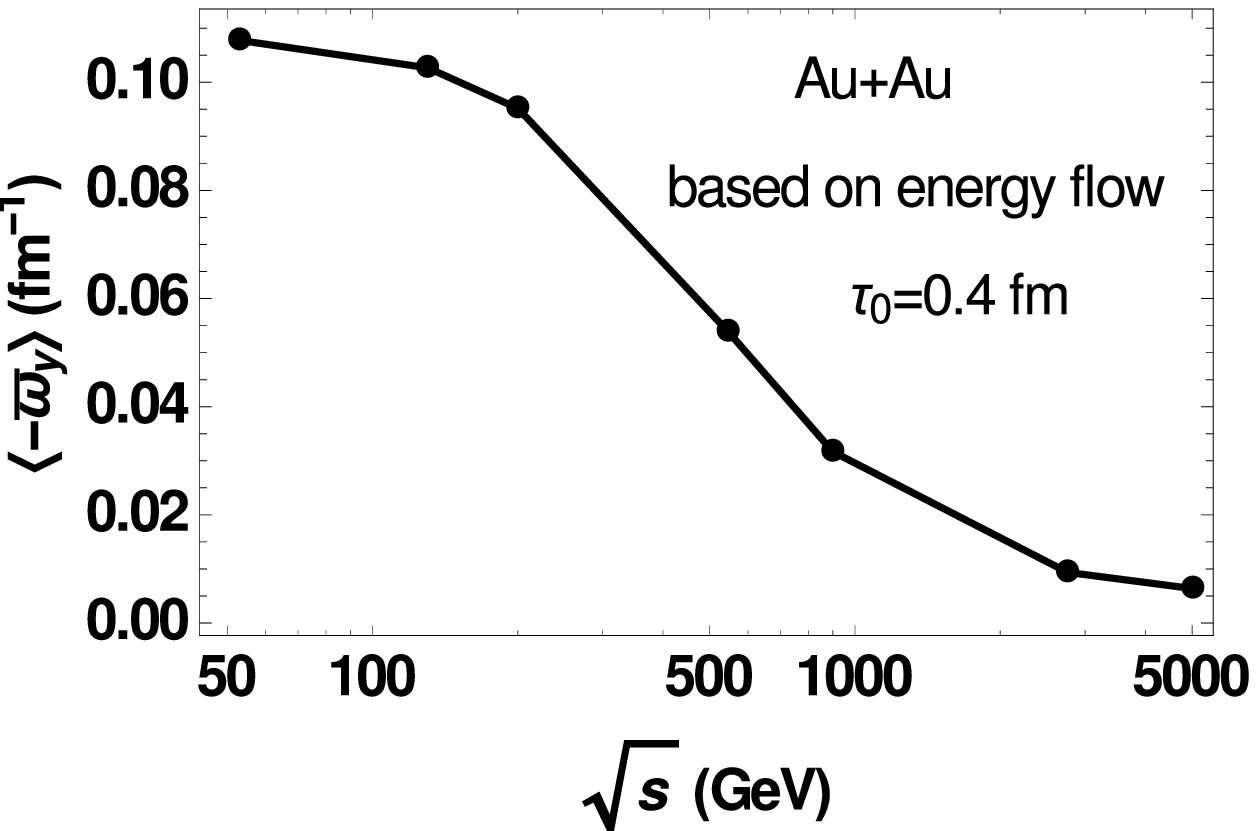}
\includegraphics[width=6cm]{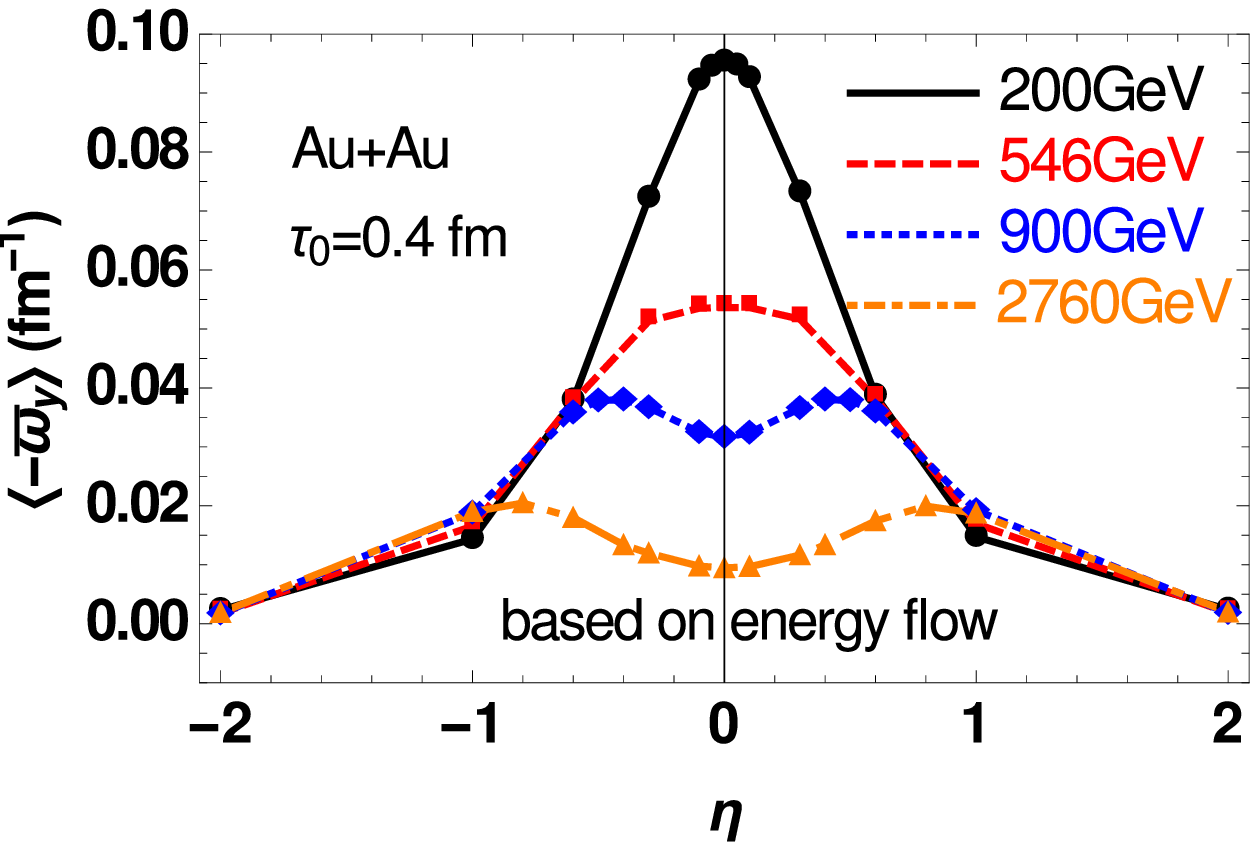}
\caption{The collision energy dependence of the vorticity at $\w=0$ (Left) and the spacetime rapidity dependence (Right) at various collision energies. Proper time is fixed $\t=0.4$ fm and impact parameter is $b=10$ fm.}
\label{fig-vor-ene}
\end{center}
\end{figure}
%%%%%%%%%%%%%%%%%%%%%%%%%%%%%%%%%%%%%%%%%%%%%%%%%%%%%%%%%%%%%%%%%%%%%%%
In \fig{fig-vor-ene} (Left) we show $\lan\bar{\o}_y\ran$  at mid-rapidity as a function of collision energy $\sqrt{s}$. Clearly, the magnitude of $\lan\bar{\o}_y\ran$ decreases when $\sqrt{s}$ increases. This, at first sight, may seem counter-intuitive as the angular momentum increases with $\sqrt{s}$. However, with increasing $\sqrt{s}$, the moment of inertia grows more rapidly than the increasing of the total angular momentum of QGP, and can make the vorticity decrease. More importantly, with increasing collision energy, more angular momentum is carried by particles at finite rapidity and thus the vorticity at $\w=0$ is relatively weakened (see \fig{fig-vor-ene} (Right)). This reflects the fact that at higher collision energy, the system at the mid-rapidity region behaves closer to the Bjorken boost invariant picture and thus allows smaller vorticity.

\begin{figure}[!htb]
\begin{center}
\includegraphics[width=5.5cm]{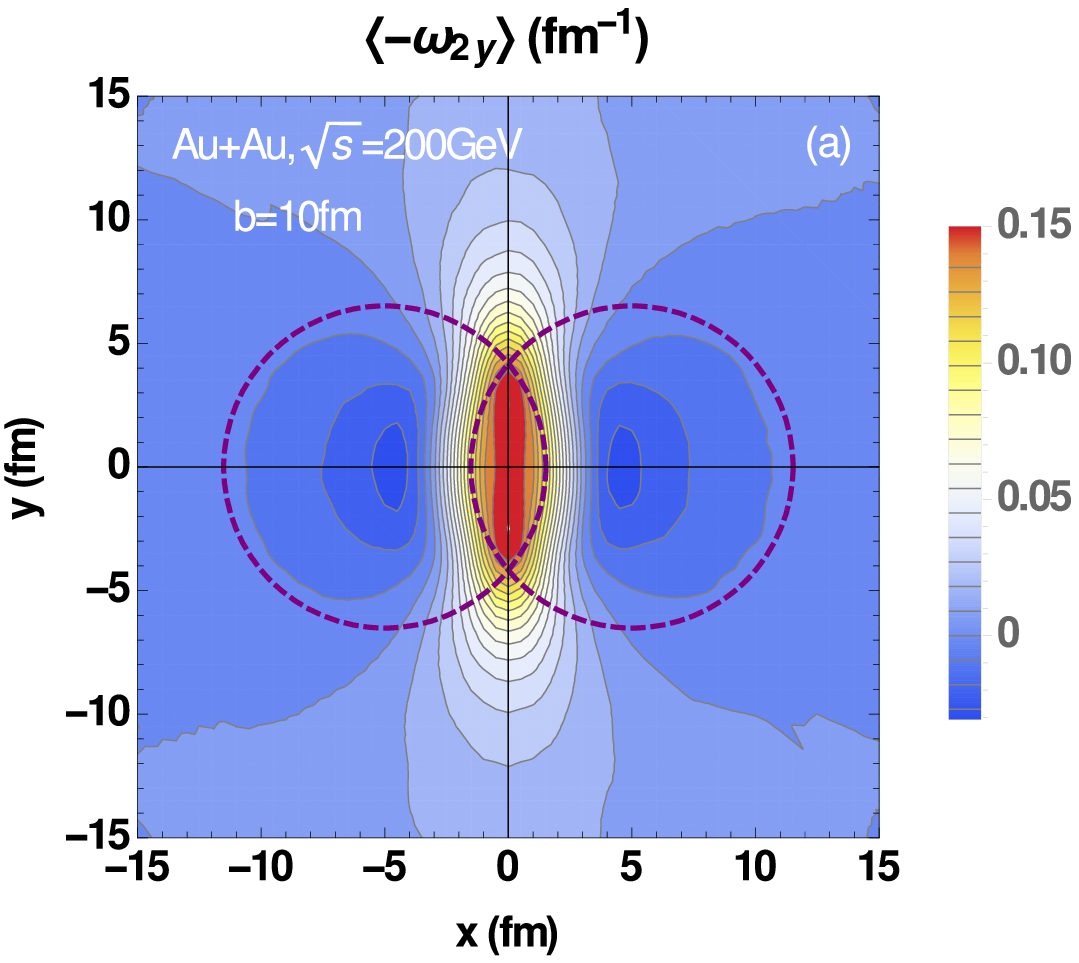}
\includegraphics[width=5.5cm]{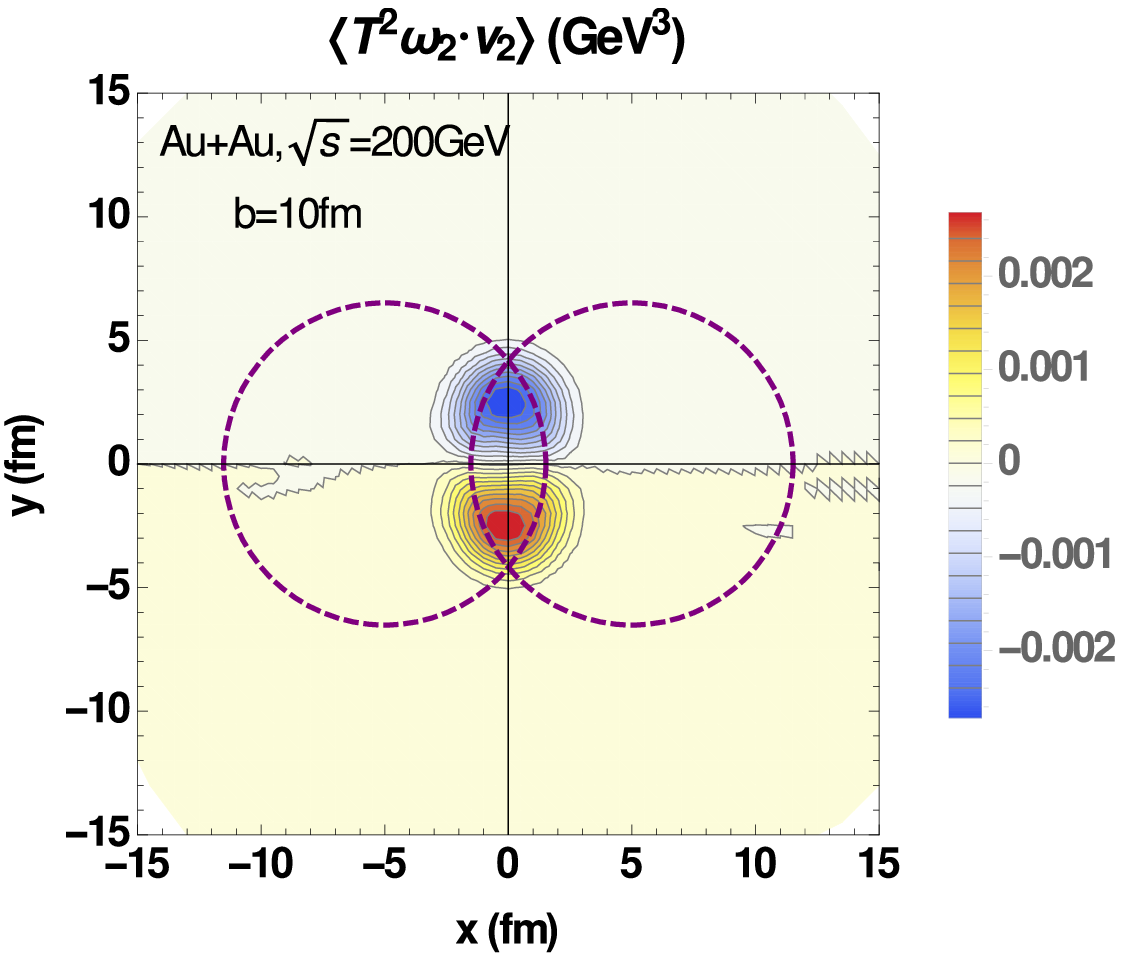}
\caption{The spatial distributions of $\lan\o_{2y}\ran$ (Left) and the event-averaged helicity, $\lan T^2\vec\o_2\cdot\bv_2\ran$ (Right), in the transverse plane for RHIC Au + Au collisions at $\sqrt{s}=200$ GeV.}
\label{fig-vor-spa-200}
\end{center}
\end{figure}
%%%%%%%%%%%%%%%%%%%%%%%%%%%%%%%%%%%%%%%%%%%%%%%%%%%%%%%%%%%%%%%%%%%%%%%
The spatial distribution of the vorticity (we present only $\lan\o_{2y}\ran$ of $\bv_2$ as an example) in the transverse plane is shown in \fig{fig-vor-spa-200} (Left). Notice that $\lan\o_{2y}\ran$ varies more steeply along the $x$ direction than along the $y$ direction in consistence with the elliptic shape of the overlapping region. The spatial distribution of the $T^2$-weighted flow helicity in the transverse plane is shown in \fig{fig-vor-spa-200} (Right). Clearly, the reaction plane separates the region with positive helicity from the region with negative helicity. The flow helicity separation may have interesting experimental implication, for example, it may be related to the chiral charges separation via the CVE~\cite{Avdoshkin:2014gpa,Yamamoto:2015gzz}.

\section{Summary}
In summary, we have performed an event-by-event study of a number of interesting features of the vorticity generated in heavy-ion collisions, including its strength, beam energy dependence, centrality dependence, and spatial distribution. More results can be found in Ref.~\cite{Deng:2016gyh}. Our study may be useful to the numerical simulation of CVE, CVW, and $\L$ baryon spin polarization.

{\bf Acknowledgments:}
W.-T.D is supported by the Independent Innovation Research Foundation of Huazhong University of Science and Technology (Grant No. 2014QN190) and the NSFC with Grant No. 11405066. X.-G.H. is supported by NSFC with Grant No. 11535012 and the One Thousand Young Talents Program of China.

\end{document}